%% file: main.tex
\documentclass[runningheads,hidelinks]{llncs}
\usepackage{amsfonts}
\usepackage{amsmath}
\usepackage[T1]{fontenc}
\usepackage{graphicx}
\usepackage{hyperref}
\usepackage{proof}
\usepackage{xspace}
\usepackage{todonotes}
\usepackage{listings}
\usepackage{tikz}
\usetikzlibrary{decorations.pathreplacing,arrows.meta}
\usepackage{vampire_induction}
\usepackage{cleveref}
\usepackage{verbatim}
\usepackage{bbding}

\usepackage[firstpage]{draftwatermark} %

\renewcommand{\paragraph}[1]{\par\smallskip\noindent\textbf{#1}}

\def\alasca{\textsc{Alasca}\xspace}

\newcommand{\CC}{C\nolinebreak\hspace{-.05em}\raisebox{.4ex}{\tiny\bf +}\nolinebreak\hspace{-.10em}\raisebox{.4ex}{\tiny\bf +}}
\DeclareMathOperator{\concat}{+\kern -0.4em+}

\newcommand\Cons[3]{#2\,\#\,#3}
\newcommand\PCons[3]{(\Cons{#1}{#2}{#3})}
\newcommand\Real{\R}

\def\orcidID#1{\href{http://orcid.org/#1}{\raisebox{-1.25pt}{\includegraphics{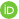}}}}

\title{The \vampire{} Diary\thanks{This is the version of the paper accepted for CAV 2025.}}
\author{
Filip~B\'artek\inst{1}\orcidID{0000-0002-1822-2651}
\and
Ahmed~Bhayat\inst{4}\orcidID{0000-0002-1343-5084}
\and
Robin~Coutelier\inst{3}\orcidID{0009-0002-4735-5215}\and 
M\'arton~Hajdu\inst{3}\orcidID{0000-0002-8273-2613}
\and
Matthias~Hetzenberger\inst{3}\orcidID{0000-0002-2052-8772}
\and
Petra~Hozzov\'a\inst{1}\orcidID{0000-0003-0845-5811}\and
Laura~Kov\'acs\inst{3}\textsuperscript{(\Envelope)}\orcidID{0000-0002-8299-2714}
\and
Jakob~Rath\inst{3}\orcidID{0000-0003-0346-6749}
\and
Michael~Rawson\inst{5}\textsuperscript{(\Envelope)}\orcidID{0000-0001-7834-1567}\and 
Giles~Reger\inst{4}\orcidID{0000-0001-6353-952X}
\and Martin~Suda\inst{1}\textsuperscript{(\Envelope)}\orcidID{0000-0001-6990-8699}\and Johannes~Schoisswohl\inst{3}\orcidID{0000-0001-5550-196X}
\and
Andrei~Voronkov\inst{2,4}\textsuperscript{(\Envelope)}\orcidID{0000-0003-1073-7615}
}

\authorrunning{The \vampire Team}

\institute{Czech Technical University in Prague, Czech Republic\\
\email{martin.suda@cvut.cz} \and EasyChair \and TU Wien, Vienna, Austria\\\email{laura.kovacs@tuwien.ac.at} \and University of Manchester, UK\\
\email{andrei@voronkov.com}
\and University of Southampton, UK\\
\email{michael@rawsons.uk} }

\SetWatermarkAngle{0}
\SetWatermarkText{\raisebox{10cm}{
\hspace{-.25cm}
\href{https://doi.org/10.5281/zenodo.15207788}{\includegraphics{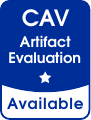}}
\hspace{8cm}
\href{https://doi.org/10.5281/zenodo.15207788}
{\includegraphics{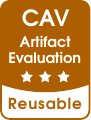}}
}}

\begin{document}
\maketitle
\begin{abstract}
During the past decade of continuous development, the  theorem prover 
\vampire has become
an automated solver for the combined theories of commonly-used data structures. \vampire{} now supports arithmetic, induction, and higher-order logic. These advances have been made to meet the demands of software verification, enabling \vampire{} to effectively complement SAT/SMT solvers and aid proof assistants. We explain how best to use \vampire in practice and review the main changes \vampire{} has undergone since its last tool presentation, focusing on the engineering principles and design choices we made during this process. 
\end{abstract}

\input{1_introduction}
\input{2_user_guide}
\input{3_demo}
\input{4_new_capabilities}
\input{5_making_it_work}

\section{Related Work and Conclusion}
We have explained how best to use \vampire, discussed new features of \vampire that  better align saturation-based first-order theorem proving with software verification, and described engineering required to make it work in practice.

As a first-order theorem prover with support for theories, induction and higher-order logic, \vampire has been influenced by, competes with, and might be variously compared to: SMT solvers such as \textsf{cvc5}~\cite{cvc5} or Z3~\cite{Z3}; first-order ATPs such as E~\cite{E19}, \textsc{Spass}~\cite{Spass09}, iProver~\cite{iProver}, or Twee~\cite{twee}; inductive theorem provers such as ACL2~\cite{ACL2book}, HipSpec~\cite{HipSpec}, or Zeno~\cite{Zeno}; and higher-order ATPs such as Zipperposition~\cite{zipperposition} or Leo-III~\cite{leo3}.
\vampire{} distinguishes itself with its native support for quantifiers combined with calculus extensions to reason about theories, induction and higher-order logic, all tied together by highly-efficient adaptive data structures and algorithms.
Naturally, \vampire{} integrates SAT and SMT solving for ground reasoning tasks. 

This paper overviewed the main reasoning engines and practices \vampire{} offers in order to assist users in understanding the many ways  \vampire{} can be integrated in other technologies. The system is under continuous development, with new applications towards proof checking and extracting system code from formal proofs.
Further advances in creating tailored \vampire{} proof schedules for proof assistants, for example in Isabelle's \emph{Sledgehammer}~\cite{sledgehammer}, are also under active development. 

\subsubsection*{Acknowledgements.} We would like to thank all users and prior developers who contributed to \vampire. We acknowledge the valuable \vampire contributions made by Daneshvar Amrollahi, Ioan Dragan, Bernhard Gleiss, Bernhard Kragl, Kry\v{s}tof Hoder, Evgenii Kotelnikov, Alexandre Riazanov, Martin Riener, Simon Robillard, Boris Shminke, and Eva Maria Wagner.

This research was funded in whole or in part by the  ERC Consolidator Grant ARTIST 101002685, the ERC Proof of Concept Grant LEARN 101213411, the TU Wien Doctoral College SecInt, the FWF SpyCoDe Grant 10.55776/F85,  the WWTF grant [ForSmart Grant ID: 10.47379/ICT22007], and the Amazon Research Award 2023 QuAT. Martin Suda was supported by the project CORESENSE no. 101070254 under the Horizon Europe programme and by the Czech Ministry of Education, Youth and Sports under the ERC CZ project POSTMAN no. LL1902.
Petra Hozzová was supported by the European Union under the project ROBOPROX (reg. no. CZ.02.01.01/00/22\_008/0004590).

\subsubsection{Disclosure of Interests.}
The authors have no competing interests to declare that are relevant to the content of this article.

\bibliographystyle{splncs04}
\bibliography{bibliography}

\end{document}

%% file: 1_introduction.tex
\section{Introduction}
Automated reasoning has become indispensable for certifying the correctness of software systems and services~\cite{DBLP:conf/cav/Rungta22}, from Boolean satisfiability (SAT) through satisfiability modulo theories (SMT) to automated theorem proving (ATP) in first-order and higher-order logic. This tool paper describes major developments in saturation-based theorem proving,
bringing our \vampire{} system to bear on modern software certification.
\vampire{} now reasons efficiently in a polymorphic first-order logic with theories, induction and quantifiers, which is realized through 
(i) combining satisfiability solving with first-order theorem proving using the AVATAR framework \cite{DBLP:conf/cav/Voronkov14,DBLP:conf/gcai/RegerB0V16}; 
(ii) native support for quantified reasoning with mixed arithmetic using extensions of superposition with quantifier elimination~\cite{ALASCA,DBLP:conf/lpar/SchoisswohlKK24};
and (iii)
embedding second-order induction schemata as inference rules in proof search~\cite{GettingSaturated22,IJCAR24InductionInvited}.
Furthermore, (iv) \vampire has evolved to support higher-order logic~\cite{DBLP:conf/cade/BhayatSR23}, program synthesis~\cite{IJCAR24}, and finding counterexamples~\cite{DBLP:conf/sat/Reger0V16}. %

Our advances in
saturation-based reasoning \emph{proved 
to make a difference}. 
\vampire outperforms or complements many other state-of-the-art reasoners, including leading SMT solvers and inductive theorem provers. For example, in CASC-J12, the most recent world championship in theorem proving, \vampire proved more problems than any other system in every competition division~\cite{CASC-J12}. %
{We believe that the increasing demand on efficient reasoning with   quantifiers, theories and induction turns \vampire into a powerful solver in the automation of  mathematics~\cite{sledgehammer}, verification of logic programs~\cite{CLINGO},  
ensuring system security~\cite{CryptoVampire}, and many other areas.
}

\begin{figure}[t]
    \centering

    \newcommand\timelineitem[3]{%
    \draw[thick,draw=gray] (#1, 0.056) to[out=180,in=-90,looseness=1.3]++ (-.5, .45) node[above,anchor=west,rotate=50] {\scriptsize \begin{tabular}{l}#3\end{tabular}};%
    \node[rotate=50,xshift=-1em,anchor=east] at (#1, .1) {\scriptsize #2};%
    }%

    \begin{tikzpicture}[outer sep=0pt,scale=.8]
        \draw[-{Triangle[width=12pt,length=8pt]},line width=8,draw=gray] (0,0) -- (13,0);
        \timelineitem{1}{2013}{First-Order Theorem\\Proving and \vampire{}}
        \timelineitem{2}{2014}{AVATAR}
        \timelineitem{2.7}{2015}{Finite model building}
        \timelineitem{3.5}{2016}{FOOL}
        \timelineitem{4.4}{2017}{Inductively defined\\datatypes}
        \timelineitem{5.5}{2018}{Unification with\\abstraction}
        \timelineitem{6.7}{2019}{Combinatory\\higher-order logic}
        \timelineitem{7.7}{2019/20}{Induction}
        \timelineitem{8.7}{2020}{Open-source under\\BSD license}
        \timelineitem{10}{2023}{ALASCA}
        \timelineitem{11}{2023}{Higher-order logic}
        \timelineitem{12}{2024}{Recursive\\synthesis}
    \end{tikzpicture}
    \caption{\vampire{} timeline since our 2013 tutorial tool paper~\cite{CAV13}.}
    \label{fig:diary}
\end{figure}
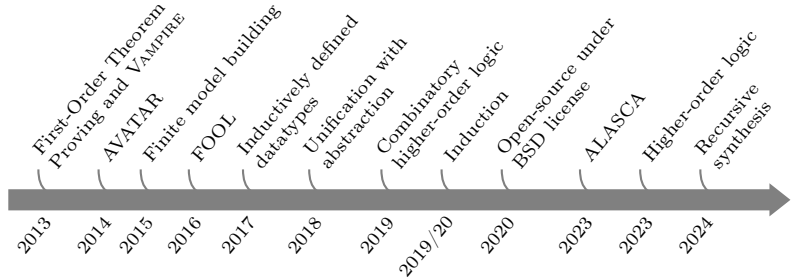

This paper details the aforementioned advances in \vampire{}.
It aims to explain how to use \vampire{} ({\Cref{sec:user-guide}}) and to give an overview of the reasoning techniques used under the hood in order to allow for reasoning in more expressive logics (\Cref{sec:features}) and more efficient reasoning in general (\Cref{sec:making-it-work}).
With these new features, a new permissive license, and unprecedented performance, we believe that a tool demonstration after more than a decade of continuous development is overdue.
Our diary of improvements since our 2013 tutorial paper  demonstration~\cite{CAV13} is summarized in \Cref{fig:diary}. 
The present paper serves as a self-contained tool demonstration, describing the many new features \vampire supports and users can exploit.
Our paper contains typical usage guidance, properly instructing readers/users interested in \vampire.
For details on \vampire's calculus, inferences processes and proof search algorithms we refer to~\cite{CAV13}.

%% file: 2_user_guide.tex
\section{User Guide}
\label{sec:user-guide}
\vampire ingests an input problem in either the TPTP~\cite{TPTP} or SMT-LIB 2~\cite{SMTLIB} formats. It then attempts to show unsatisfiability\footnote{TPTP allows the user to supply a \texttt{conjecture}: this formula is automatically negated.} of the input by deriving falsum through application of its logical calculus.
If it succeeds, \vampire{} halts and prints a step-by-step refutation.
In some cases \vampire can also show satisfiability. This happens either by finding a finite model (Section~\ref{sec:sat-and-smt}) or by \emph{saturation}: detecting that a refutation cannot be derived with a complete calculus.

\paragraph{Licensing.}
\vampire{} is now open-source and available online\footnote{\url{https://github.com/vprover/vampire/}} under a 3-clause BSD license.
This has completely changed the Vampire team dynamics and coding culture: the development team grew, code became fully shared, and external contributors can improve \vampire.
The new license has increased the user base of Vampire in teaching, research and development.

\vampire uses external code for specific tasks, including the \textsc{MiniSat}~\cite{DBLP:conf/sat/EenS03} and \textsc{CaDiCaL}~\cite{cadical} SAT solvers, the Z3 SMT solver~\cite{Z3} (optional), the VIRAS quantifier elimination routine~\cite{DBLP:conf/lpar/SchoisswohlKK24}, and \texttt{mini-gmp}~\cite{gmp} for arbitrary-precision arithmetic.

\paragraph{Installation.}
\vampire{} is written in \CC17~\cite{cpp17} and uses the CMake~\cite{cmake} build system.
Since the previous tool demonstration over a decade ago~\cite{CAV13}, a series of patches have improved portability and \vampire{} now runs on a variety of modern architectures and UNIX-like operating systems.
We provide pre-compiled binaries for UNIX-like systems, and it is generally straightforward to compile \vampire{} from source on such systems.
Support for other operating systems is more experimental, but users report success with compatibility layers such as the Windows Subsystem for Linux~\cite{WSL}, Cygwin~\cite{cygwin}, or Cosmopolitan Libc~\cite{cosmopolitan}.

\paragraph{Quick Start.}
It is possible, but \emph{usually undesirable}, to invoke \vampire directly on an input file.
This will cause it to run a single proof attempt, which will likely not be well-suited to the input.
Instead, users should \emph{schedule} an execution of a \emph{portfolio} of many different \emph{strategies}, which is achieved with the invocation
\[\texttt{vampire {-}{-}mode portfolio {-}{-}schedule <schedule> {-}{-}cores 0 <problem>}\]
where \texttt{<problem>} is the input problem, \texttt{{-}{-}cores 0} instructs \vampire to use all available CPU cores, and \texttt{<schedule>} should be selected from \vampire's list of built-in schedules (Section~\ref{sec:portfolio}), depending on the input. %

\paragraph{Understanding the Output.}
\vampire prints status messages as new strategies are launched or old strategies fail, until either a strategy succeeds or the time limit is reached.
When a strategy succeeds, by default \vampire reports an SZS status~\cite{DBLP:conf/lpar/Sutcliffe08} and then, assuming that the input is unsatisfiable, a human-readable proof.
The most common SZS statuses \vampire reports are:
\begin{description}
    \item[\texttt{Satisfiable}:] input does not contain a conjecture and is satisfiable 
    \item[\texttt{Unsatisfiable}:] input does not contain a conjecture and is unsatisfiable
    \item[\texttt{CounterSatisfiable}:] conjecture present, after negation the input is satisfiable
    \item[\texttt{Theorem}:] conjecture present, after negation the input is unsatisfiable
    \item[\texttt{ContradictoryAxioms}:] conjecture present, but axioms alone are unsatisfiable
\end{description}

\paragraph{Controlling Output.}
\vampire by default uses the SZS standards for output and reports a compact, human-readable proof.
The \vampire{} output can be changed with the \emph{output mode} (\texttt{-om}) and \emph{proof format} (\texttt{-p}) flags.
For example, \texttt{-om smtcomp} produces a very terse output suitable for %
SMT-COMP~\cite{SMTComp}, while \texttt{-p tptp} produces machine-readable TSTP~\cite{DBLP:conf/lpar/Sutcliffe10} proofs.
Some more exotic formats such as \LaTeX{} (\texttt{{-}{-}latex\_output}) or Dedukti~\cite{dedukti} are under development.

\paragraph{Setting Resource Limits.} \vampire can be configured to limit the amount of time (\texttt{-t <seconds>}), memory (\texttt{-m <MB>}), and on recent Linux systems the number of userspace instructions retired (\texttt{-i <MI>}\footnote{Millions of instructions~\cite{DBLP:conf/cade/Suda22}. Instruction limits tend to more stable than time limits across hardware and operating system conditions.}). The value \texttt{0} means no limit.
\paragraph{Exploring Options.}
\vampire has many, many options. 
A complete list can be generated with \texttt{{-}{-}show\_options on}, and a particular flag can be examined with \texttt{{-}{-}explain}, e.g. \texttt{{-}{-}explain output\_mode}.
Most options only affect a single strategy's behavior, but some affect the global behavior of \vampire, such as the proof format or the time limit.
Single strategy options are not usually controlled by the user but automatically set by portfolio schedules (more on this in \Cref{sec:portfolio}).

\subsection{Looking Under the Hood}
It is sometimes useful for users to inspect the internal state of \vampire, such as when debugging or optimizing an encoding.
Here we sketch the internal mechanisms of \vampire and explain how to inspect them during operation.

\vampire works in two phases. First, the input is parsed, the conjecture --- if present --- is negated, and the resulting formulas are converted to clause normal form (CNF) and preprocessed. 
Then \vampire tries to derive the empty clause (witnessing contradiction) in order to refute the CNF of the input problem.
New clauses are derived from old by applying generating inferences in \vampire's proof calculus, \emph{superposition}~\cite{CAV13}.
The search space is partitioned into three sets of clauses: \emph{new} clauses have been freshly derived; \emph{passive} clauses survived \vampire's simplification efforts but have not yet participated in inference, and \emph{active} clauses have themselves participated in inferences generating new clauses.
\vampire allows inspection of these processes. To show the clauses resulting from an input, use 
\[
\texttt{vampire {-}{-}mode clausify <problem>}
\]
which causes \vampire to stop after preprocessing {\tt <problem>} and print the resulting CNF. %
The CNF may be surprising at times as \vampire's preprocessing will happily eliminate parts of the input that it can detect will not help reaching a refutation. To inspect the progress of the preprocessing pipeline users may \texttt{{-}{-}show\_preprocessing on}. 
Once proof search begins,  \texttt{{-}{-}show\_new on} displays new clauses (analogously, \texttt{show\_passive} and \texttt{show\_active}).
Not all new clauses will make it to \emph{passive}: use \texttt{{-}{-}show\_reductions on} to see the simplifications that \vampire applies.

%% file: 3_demo.tex
\section{Demonstration: Arithmetic and Induction}
Consider the proposition ``the sum of two lists of real numbers is equal to the sum of their concatenation''. While clearly true, a formal proof requires reasoning about arithmetic, algebraic datatypes, recursive functions, polymorphism, and at least one inductive step.
When given the natural first-order formalization of the problem (Fig.~\ref{fig:motivating:fol}) encoded into version 2.7 of SMT-LIB, \vampire{} is able to find a proof immediately.
We show a distilled version of the proof in mathematical notation in Figure~\ref{fig:proof:fol}.

\newcommand\Generic[1]{\langle#1\rangle}
\newcommand\sumf{\textrm{sum}}
\newcommand\R{\mathbb{R}}
\begin{figure}
    \[
      \begin{array}{rrl}
        &\sumf(\epsilon) &= 0\\
        \forall x: \R.~\forall xs: [\R].& \sumf(\Cons{\Real}{x}{xs}) &= x + \sumf(xs) \\
        \Lambda \alpha.~\forall ys: [\alpha].& \epsilon \concat ys &= ys\\
        \Lambda \alpha.~\forall x:\alpha.~\forall xs, ys : [\alpha].& (\Cons{\alpha}{x}{xs}) \concat ys &= \Cons{\alpha}{x}{(xs \concat ys)}\\
        \hline
        \forall xs,ys:[\R]. &\sumf(xs) + \sumf(ys) &= \sumf(xs \concat ys)
      \end{array}
    \]
    \caption{Motivating example in polymorphic first-order logic with uninterpreted functions $\textrm{sum} : [\R] \to \R$, $\concat : \Lambda \alpha.~[\alpha] \times [\alpha] \to [\alpha]$, algebraic datatypes and real arithmetic.}
    \label{fig:motivating:fol}
\end{figure}
\begin{figure}
    \input{proof.tex}
    \caption{\vampire's proof output of the problem from Fig.~\ref{fig:motivating:fol} in mathematical notation.
    The symbols $\sK i$ are fresh Skolem constants.
    The induction steps are detailed in Sect.~\ref{sec:features:induction}.} 
    \label{fig:proof:fol}
\end{figure}

The proof displays some new features of \vampire, in particular the use of structural induction (Sect.~\ref{sec:features:induction}) and superposition-based arithmetic reasoning via the ALASCA calculus %
(Sect.~\ref{sec:features:arithmetic}). These features are key to \vampire's success on this problem. 
The need for our unique blend of induction, arithmetic and polymorphism is supported by the fact that other solvers such as CVC5~\cite{cvc5} or Z3~\cite{Z3} cannot, to the best of our knowledge, yet process or prove problems such as this one.

%% file: proof.tex
\newcommand\ProofLine[4]{#1. & #3 & \textsf{#4} \\}
\newcommand\Concat[3]{#2 \concat #3}
\newcommand\PConcat[3]{(\Concat{#1}{#2}{#3})}
\newcommand\Nil[1]{\epsilon}
\newcommand\sK[1]{\sigma_{#1}}

\[
\begin{array}{rll}
\ProofLine{1}{13}{ 0 = \sumf(\Nil{ \Real }) }{ assumption 1 (cnf) }
\ProofLine{2}{14}{ 0 = x + \sumf(y) - \sumf(\Cons{\Real}{x}{y}) }{ assumption 2 (cnf) }
\ProofLine{3}{15}{\Concat{\alpha}{\Nil{ \alpha }}{x} = x }{ assumption 3 (cnf) }
\ProofLine{4}{16}{\Cons{\alpha}{y}{\PConcat{\alpha}{z}{x}} = \Concat{\alpha}{\PCons{\alpha}{y}{z}}{x} }{ assumption 4 (cnf) }
\ProofLine{5}{17}{ 0 \neq \sumf(\sK1) + \sumf(\sK0) - \sumf(\Concat{\Real}{\sK1}{\sK0}) }{ conjecture (cnf) }

\ProofLine{6}{24}{ 
  \begin{array}[t]{rl}
       & 0 \neq \sumf(\Cons{\Real}{\sK3}{\sK4}) + \sumf(\sK0) - \sumf(\Concat{\Real}{\PCons{\Real}{\sK3}{\sK4}}{\sK0})  \\
  \lor & 0 \neq \sumf(\Nil{ \Real }) + \sumf(\sK0) - \sumf(\Concat{\Real}{\Nil{ \Real }}{\sK0}) 
  \end{array}
  }{ structural induction 5 }%
\ProofLine{7}{25}{ 
  \begin{array}[t]{rl}
       & 0 = \sumf(\sK4) + \sumf(\sK0) - \sumf(\Concat{\Real}{\sK4}{\sK0}) \\
  \lor & 0 \neq \sumf(\Nil{ \Real }) + \sumf(\sK0) - \sumf(\Concat{\Real}{\Nil{ \Real }}{\sK0}) \\
  \end{array}
  }{ structural induction 5 } %
\ProofLine{8}{34}{ 
  \begin{array}[t]{rl}
         & 0 \neq \sumf(\sK0) - \sumf(\Cons{\Real}{\sK3}{\PConcat{\Real}{\sK4}{\sK0}}) + \sumf(\Cons{\Real}{\sK3}{\sK4}) \\
    \lor & 0 \neq \sumf(\sK0) + \sumf(\Nil{ \Real }) - \sumf(\Concat{\Real}{\Nil{ \Real }}{\sK0}) \\
  \end{array}
  }{ forward demodulation 6,4 }
\ProofLine{9}{36}{ 
  \begin{array}[t]{rl}
         & 0 \neq \sumf(\sK0) + \sumf(\Nil{ \Real }) - \sumf(\sK0) \\
    \lor & 0 = \sumf(\sK0) - \sumf(\Concat{\Real}{\sK4}{\sK0}) + \sumf(\sK4) \\
  \end{array}
  }{ forward demodulation 7,3 }
\ProofLine{10}{37}{ 
  \begin{array}[t]{rl}
       & 0 \neq \sumf(\Nil{ \Real }) \\
  \lor & 0 = \sumf(\sK0) - \sumf(\Concat{\Real}{\sK4}{\sK0}) + \sumf(\sK4) \\
  \end{array}
  }{ \alasca normalization 9 }
\ProofLine{11}{42}{ 
  \begin{array}[t]{rl}
         & 0 \neq \sumf(\sK0) + \sumf(\Nil{ \Real }) - \sumf(\sK0)\\
    \lor & 0 \neq \sumf(\sK0) + \sumf(\Cons{\Real}{\sK3}{\sK4}) - \sumf(\Cons{\Real}{\sK3}{\PConcat{\Real}{\sK4}{\sK0}}) \\
  \end{array}
  }{ forward demodulation 8,3 }
\ProofLine{12}{43}{ 
  \begin{array}[t]{rl}
         & 0 \neq \sumf(\Nil{ \Real }) \\
    \lor & 0 \neq \sumf(\sK0) + \sumf(\Cons{\Real}{\sK3}{\sK4}) - \sumf(\Cons{\Real}{\sK3}{\PConcat{\Real}{\sK4}{\sK0}}) \\
  \end{array}
}{ \alasca normalization 11 }
\ProofLine{13}{44}{ 0 = \sumf(\sK0) - \sumf(\Concat{\Real}{\sK4}{\sK0}) + \sumf(\sK4) }{ subsumption resolution 10,1 }
\ProofLine{14}{59}{ 0 \neq \sumf(\sK0) + \sumf(\Cons{\Real}{\sK3}{\sK4}) - \sumf(\Cons{\Real}{\sK3}{\PConcat{\Real}{\sK4}{\sK0}}) }{ subsumption resolution 12,1 }
\ProofLine{15}{84}{ 0 \neq \sumf(\sK0) + \sumf(\Cons{\Real}{\sK3}{\sK4}) - (\sK3 + \sumf(\Concat{\Real}{\sK4}{\sK0})) }{ \alasca superposition 2,14 }
\ProofLine{16}{262}{ \thickmuskip=3mu \medmuskip=2mu \thinmuskip=1mu 0 \neq \sumf(\sK0) - \sK3 + \sumf(\Cons{\Real}{\sK3}{\sK4}) - (\sumf(\sK0) + \sumf(\sK4)) }{ \alasca superposition 13,15 }
\ProofLine{17}{275}{ 0 \neq \sK3 - \sumf(\Cons{\Real}{\sK3}{\sK4}) + \sumf(\sK4) }{ \alasca normalization 16 }
\ProofLine{18}{312}{ 0 \neq \sK3 - (\sK3 + \sumf(\sK4)) + \sumf(\sK4) }{ \alasca superposition 2,17 }
\ProofLine{19}{319}{ \Box }{ \alasca normalization 18 }
\end{array}
\]

%% file: 4_new_capabilities.tex
\section{New Capabilities} 
\label{sec:features}
Here we present the most significant new capabilities implemented in \vampire{} since 2013~\cite{CAV13}. Improvements to existing capabilities are in Section~\ref{sec:making-it-work}.

\subsection{Arithmetic Reasoning}
\label{sec:features:arithmetic}
Reasoning about arithmetic in the presence of quantifiers is highly desirable. 
To this end, \vampire{} implements the Abstracting Linear Arithmetic Superposition Calculus (\alasca) \cite{ALASCA} that combines ideas like inequality chaining, unification with abstraction~\cite{Reger0V18}, and rewriting modulo linear arithmetic. %
Equality reasoning in \alasca can be seen as applying the superposition calculus modulo the axioms of linear arithmetic. For example, take step 18 of \Cref{fig:proof:fol}. There, the literal $0 = x + \sumf(y) - \sumf(\Cons{\R}{x}{y})$ is used to perform a rewrite $\sumf(\Cons{\R}{x}{y}) \leadsto x + \sumf(y)$ instead of a rewrite $x + \sumf(y) - \sumf(\Cons{\R}{x}{y}) \leadsto 0$, which would be the only permissible rewrite in standard superposition. 
In addition, \alasca uses inequality chaining and dedicated factoring rules to deal with inequalities, and combines variable elimination rules with unification with abstraction to efficiently perform unification, modulo linear arithmetic.

\paragraph{Non-Linear Reasoning.}
\alasca itself supports reasoning in linear real arithmetic with uninterpreted functions and quantifiers. Nonlinear problems are also supported by treating nonlinear multiplications as uninterpreted functions, automatically adding the relevant axiomatization.

\paragraph{Mixed Integer-Real Arithmetic.}
While the original \alasca work is limited to real arithmetic, our current implementation in \vampire{} lifts these restrictions. We support reasoning in mixed integer-real arithmetic, using a tailored quantifier-elimination procedure~\cite{DBLP:conf/lpar/SchoisswohlKK24}, as well as various new inference rules\footnote{The work on this inference system has not been published yet.} to handle the combination of mixed arithmetic and uninterpreted functions by natively supporting the rounding (floor) function.

\paragraph{Simplifications and Generalizations.}
In addition to \alasca{}, \vampire also provides  lightweight arithmetic reasoning~\cite{TheorySimplifications}. This includes arithmetic subterm generalization rules that complement \alasca reasoning, and other simplification rules entailed by \alasca. 
Although these simplification rules are not as widely applicable as \alasca, they provide for more lightweight and therefore efficient arithmetic reasoning, sufficient for many practical problems.
The generalization rules include transformations like turning $\forall x,y : \R.~P(3 x + y) $ into the equivalent clause $ \forall x : \R.~P(x)$.

\paragraph{Integrating SMT Solvers.}
\vampire sometimes hands off \emph{ground}, that is quantifier-free, arithmetic reasoning to the Z3 SMT solver~\cite{Z3}. This is done either by invoking AVATAR modulo theories~\cite{DBLP:conf/gcai/RegerB0V16} (see Sect.~\ref{sec:sat-and-smt}) or by \emph{theory instantiation} \cite{DBLP:conf/tacas/Reger0V18}.
Theory instantiation uses an SMT solver to find possible instantiations of clauses based on their purely arithmetical literals. To illustrate, consider the clause $\forall x : \mathbb{Z}.~P(x) \lor 0 > 3 x \lor x \geq 1$. The SMT solver is queried for a model satisfying $\lnot (0 > 3 x \lor x \geq 1)$, which is only the case for $x = 0$. The clause is instantiated with $\{ x \mapsto 0 \}$ and simplified to $P(0)$.
Such integration of SMT solvers enables using  state-of-the-art developments in SMT and is particularly beneficial for problem areas such as  non-linear reasoning, for which \vampire does not yet have dedicated calculi.

\subsection{Inductive Reasoning}
\label{sec:features:induction}
\vampire{} supports inductive reasoning~\cite{IJCAR24InductionInvited} over literals with up to one free variable.\footnote{This covers the cases of a universally-quantified conjecture, and a conjecture with any number of universally-quantified variables and one existentially-quantified variable.}
It applies induction by generating theory lemmas, triggered by deriving an eligible induction goal.
\vampire{} supports structural induction over inductively-defined datatypes~\cite{CADE19}, induction over bounded intervals of integers~\cite{CADE21}, and well-founded induction principles generated from recursive function definitions~\cite{FMCAD21}.
Immediately after \vampire{} generates the induction lemmas, it uses them to resolve their corresponding goals.

A distinctive feature of \vampire{} is that it seamlessly interleaves induction with other inferences, efficiently handling hundreds of thousands of induction formulas. %
This makes it possible to use more explosive lemma generation techniques essential for solving some inductive problems. To synthesize lemmas, \vampire{} can generalize over terms and term occurrences~\cite{CICM20} or over multiple literals and clauses~\cite{PAAR22}, use function definitions~\cite{FMCAD21} and perform general rewriting~\cite{LPAR24Hajdu}.

Let us highlight some key steps of the automated induction in \vampire{} using the proof from Figure~\ref{fig:proof:fol}.
First, when \vampire{} sees clause~5, it detects that induction might be in order, as clause 5 corresponds to a universally-quantified goal using an inductively-defined datatype.
Therefore, \vampire{} uses clause~5 to
instantiate the structural induction axiom for lists,
\[L[\epsilon]\land \forall x : \alpha, y : [\alpha]. (L[y]\rightarrow L[\Cons{\alpha}{x}{y}]) \rightarrow \forall z : [\alpha]. L[z],\]
by setting $\alpha:=\mathbb{R}$ and $L[t]:=0=\sumf(t)+\sumf(\sigma_0)-\sumf(t\concat \sigma_0)$. Note that $L[\sigma_1]$ is set to be complementary to the literal from clause~5.

The instantiated axiom concludes that $L[z]$ holds for any $z : [\mathbb{R}]$, while clause~5 expresses that $L[\sigma_1]$ does not hold.
To use this, \vampire{} converts the axiom into CNF, obtaining clauses $\lnot L[\epsilon]\lor L[\sigma_4]\lor L[z]$ and $\lnot L[\epsilon]\lor \lnot L[\Cons{\R}{\sigma_3}{\sigma_4}]\lor L[z]$, where $\sigma_3, \sigma_4$ are Skolem constants corresponding to $x$ and $y$, respectively.
These clauses together express that either the antecedent of the axiom does not hold (the base case $L[\epsilon]$ does not hold, or for some $\sigma_3, \sigma_4$ we have $L[\sigma_4]$ but not $L[\Cons{\R}{\sigma_3}{\sigma_4}]$), or the conclusion that $L[z]$ is true for any $z$ must hold.
Then \vampire{} applies binary resolution on these two clauses with clause~5, resolving away $L[z]$, and deriving clauses~6 and~7.
Clauses~6 and~7 are exactly $\lnot L[\epsilon]\lor L[\sigma_4]$ and $\lnot L[\epsilon]\lor \lnot L[\Cons{\R}{\sigma_3}{\sigma_4}]$, spelled out in full in  Figure~\ref{fig:proof:fol}.
The rest of the proof then covers the refutation of these two clauses.

\subsection{Polymorphic Logic}
\vampire now supports rank-1 polymorphic types~\cite{polymorphic-vampire} in the tradition of Standard ML~\cite{standard-ml}.
This represents a trade-off between expressivity and ease of implementation.
Note that \vampire does \emph{not} presently implement a sort inference routine and all sorts in non-variable terms must be explicitly given as sort arguments~\cite{tff1}, which may themselves be variables.
For example, the axiom $\sumf(\epsilon) = 0$ is actually represented as \texttt{sum(nil(\$real)) = 0}, and $\epsilon \concat ys = ys$ as \texttt{concat(A, nil(A), Ys) = Ys}.
The original motivation for introducing polymorphic logic was supporting combinatory higher-order logic~\cite{combinatory-HOL}, but it has also proved useful for supporting polymorphic theories such as arrays, and for verifying programs that use parametric polymorphism.

\subsection{Beyond First-Order Logic} %
\label{sec:features:beyond}

\newcommand\fool{\textsc{Fool}\xspace}

\vampire{} supports extensions of first-order logic useful for software analysis and verification. In particular, \vampire{} implements \fool\cite{DBLP:conf/cpp/KotelnikovKRV16}, a conservative extension of many-sorted first-order logic with \emph{if-then-else} and \emph{let-in} expressions, which can be used to capture the next-state relation of loop-free programs~\cite{DBLP:conf/cade/KotelnikovKV18}. As such, \vampire also supports first-class \emph{Boolean sorts}, by encoding the axiom $\forall x : o.~x = 0 \vee x = 1$
as a new inference rule.
The rule exploits the two-element domain property of the Boolean sort without blowing up proof search.

\paragraph{Higher-Order Logic.}
In addition, a branch of \vampire\footnote{\url{https://github.com/vprover/vampire/tree/hol}}%
implements a {super\-po\-si\-tion}-based calculus tailored for higher-order logic~\cite{DBLP:conf/ijcar/BhayatS24}, while still using the general saturation framework from first-order logic. 
As higher-order unification is undecidable, our implementation bypasses eager unification by performing bounded-depth unification and introducing constraints for remaining unification terms.
This technique of constraint introduction has also been used in pure first-order reasoning as delayed unification~\cite{DBLP:conf/cade/BhayatSR23} and in arithmetic reasoning as unification with abstraction~\cite{ALASCA,RefiningUWA}.

\subsection{Synthesis}
We further utilize \vampire{}'s powerful proving capabilities to extend it to a program synthesizer~\cite{CADE23,IJCAR24}.
\vampire{} works with a relational input-output specification expressed in first-order logic, capturing ``for all inputs 
 $x$ there exists an output $y$ such that a given relation between $x$ and $y$ holds''.
In parallel to proving this conjecture, \vampire{} constructs a program which computes the value of $y$ for any given value of $x$.
To switch on synthesis mode, use \texttt{-qa synthesis}.\footnote{While synthesis of recursion-free programs is available in the mainline \vampire{}, synthesis of recursive programs is currently in the branch \texttt{synthesis-recursive}.}

%% file: 5_making_it_work.tex
\section{Making It Work}
\label{sec:making-it-work}
Taking the above extensions into account, \vampire must now prove theorems in a substantially richer logic with a much greater number of possible inferences.
We now describe improvements to \vampire's core that we consider most important for meeting this challenge and maintaining good performance in practice. This adds to the observations of our previous tutorial paper~\cite{CAV13}, which remain valid. 
We hope to provide useful information here for those readers who develop their own reasoning systems.

\subsection{Preprocessing}
Computing normal forms and preprocessing remain of vital importance: the right normal form can eliminate much search space or drastically shorten the required proof. To this end \vampire has grown a new top-down clausal normal form routine~\cite{newcnf}, lifted the \emph{blocked clause elimination} technique~\cite{bce} from SAT~\cite{blocked-clause-elimination}, and adapted a highly effective goal-oriented rewriting technique from Twee~\cite{twee}. As a general rule of thumb, preprocessing techniques have linear-time complexity, and avoid recursion to prevent stack overflow on large inputs.

\subsection{Integrating SAT and SMT}
\label{sec:sat-and-smt}
One of the tricks for efficiently tackling real-life problems in rich formalisms such as first-order logic with theories 
is to look for sub-problems in simpler logics and offload them to dedicated tools.
In this spirit, \vampire implements the AVATAR architecture for clause splitting \cite{DBLP:conf/cav/Voronkov14}, which allows a prover to delegate the ``propositional essence'' of the given problem to a SAT solver. In AVATAR modulo theories \cite{DBLP:conf/gcai/RegerB0V16}, \vampire uses a finer abstraction\footnote{We remark that quantifiers are always handled natively by \vampire.} and delegates ground \emph{theory} sub-problems to an SMT solver. 

SAT solving is also applied within \vampire{} to find counterexamples to false conjectures. \vampire now provides a MACE-style finite model building mode, using a translation to SAT~\cite{claessen2003new,DBLP:conf/sat/Reger0V16}. This is often a useful complement to theorem-proving modes (provided that a small counter-model exists), which helps terminate futile searches early and delivers useful insights in the form of bug traces. 

\subsection{Redundancy and Proof Search}
Redundancy elimination is key to efficient proof search. Intuitively, a clause is redundant if it is a logical consequence of smaller clauses from the search space:
checking whether a first-order clause is redundant is therefore undecidable in general. \vampire{} implements cheap conditions for detecting some cases of redundancy. The central technique used to implement these checks efficiently is \emph{term indexing}~\cite{term-indexing} and here in particular \emph{substitution trees}~\cite{substitution-tree} and \emph{code trees}~\cite{PartiallyAdaptiveCodeTrees}.
Code trees are used for rewriting clauses by unit equalities~\cite{FMCAD21,LPAR24Hajdu} and eliminating duplicate clauses, while substitution trees are used for other inferences.
To efficiently solve term ordering constraints in redundancy elimination, \vampire{} uses \emph{term ordering diagrams}, which offer runtime-specialized implementations of %
simplification orderings~\cite{tod}. Finally, \vampire{} also uses SAT solving to check some redundancy conditions that can be modeled as at-most-one ground constraints over the Boolean structure of clause sets~\cite{SATSubsumption24}. %

\subsection{A Sea of Options, Strategies, and Schedules}

\label{sec:options}
\label{sec:portfolio}
By \emph{strategy} we mean a particular configuration of \vampire's option values.
Since the behavior of \vampire is controlled by more than 200 options,
the number of available strategies is vast. 
Although expert users may sometimes have an idea of which options could be well suited to tackle a problem, the prover's behavior tends to be so chaotic \cite{DBLP:conf/cade/Suda22}
that even expert hunches often fail. For this reason, \vampire provides \emph{schedules} 
of pre-selected strategies executed sequentially, possibly adapting to the given problem's features.

Creating powerful schedules is a challenging problem. Since 2010, \vampire has employed
a dedicated support tool \emph{Spider}~\cite{Spider} to construct schedules from a set of training problems. Spider trials random strategies to solve as many training problems as possible and eventually selects those strategies that \emph{complement} each other particularly well and lead to a schedule with good coverage and a short overall runtime. Techniques have recently been developed to encourage the generalization of the constructed schedule to unseen problems~\cite{DBLP:conf/ijcar/BartekCS24}.

Actively maintained schedules include \texttt{casc} and \texttt{casc\_sat}, 
for general first-order theorem proving and disproving, resp.,
referring to the famous championship \cite{Sut16}. Similarly, \texttt{smtcomp} has its origin in another competition \cite{WCDHNR19} and is optimized to work well on problems requiring theory reasoning.
The higher-order branch (Section~\ref{sec:features:beyond}) provides schedules for reasoning in higher-order logic and the \emph{Sledgehammer}~\cite{sledgehammer} use-case.
Finally, there is also \texttt{induction} and more.\footnote{Use \texttt{{-}{-}explain\_option schedule} to list schedules available from your \vampire.}%

\subsection{Branches}
Some extensions to \vampire would have violent and extensive impact on the code base.
This is true of \vampire's higher-order logic (HOL), for example.
Integrating the HOL extension into \vampire would be a significant amount of work and impose a burden on all \vampire developers: but we would like it to continue, as it is a world-leading system for higher-order logic.
The way we are currently dealing with this tension is by keeping this kind of feature on \texttt{git} branches, which are periodically synchronised with mainline \vampire.
When a branch is widely-used enough, stable, and has a clear path to be integrated cleanly with mainline \vampire, we may consider merging it: this has happened in the past with rank-1 polymorphism and a previous approach to HOL~\cite{combinatory-HOL}.

%% file: main.bbl
\begin{thebibliography}{10}
\providecommand{\url}[1]{\texttt{#1}}
\providecommand{\urlprefix}{URL }
\providecommand{\doi}[1]{https://doi.org/#1}

\bibitem{dedukti}
Assaf, A., Burel, G., Cauderlier, R., Delahaye, D., Dowek, G., Dubois, C.,
  Gilbert, F., Halmagrand, P., Hermant, O., Saillard, R.: {Dedukti: a Logical
  Framework based on the {\(\lambda\)}{\(\Pi\)}-Calculus Modulo Theory}. CoRR
  \textbf{abs/2311.07185} (2023)

\bibitem{cvc5}
Barbosa, H., Barrett, C.W., Brain, M., Kremer, G., Lachnitt, H., Mann, M.,
  Mohamed, A., Mohamed, M., Niemetz, A., N{\"{o}}tzli, A., Ozdemir, A.,
  Preiner, M., Reynolds, A., Sheng, Y., Tinelli, C., Zohar, Y.: {cvc5: {A}
  Versatile and Industrial-Strength {SMT} Solver}. In: TACAS. pp. 415--442
  (2022)

\bibitem{SMTLIB}
Barrett, C., Fontaine, P., Tinelli, C.: {The Satisfiability Modulo Theories
  Library (SMT-LIB)}. {\tt www.SMT-LIB.org} (2016)

\bibitem{SMTComp}
Barrett, C., de~Moura, L., Stump, A.: {SMT-COMP: Satisfiability modulo Theories
  Competition}. In: CAV. p. 20–23 (2005)

\bibitem{DBLP:conf/ijcar/BartekCS24}
B{\'{a}}rtek, F., Chvalovsk{\'{y}}, K., Suda, M.: {Regularization in
  Spider-Style Strategy Discovery and Schedule Construction}. In: IJCAR. pp.
  194--213 (2024)

\bibitem{RefiningUWA}
Bhayat, A., Korovin, K., Kov{\'{a}}cs, L., Schoisswohl, J.: {Refining
  Unification with Abstraction}. In: LPAR. pp. 36--47 (2023)

\bibitem{combinatory-HOL}
Bhayat, A., Reger, G.: {A Combinator-Based Superposition Calculus for
  Higher-Order Logic}. In: IJCAR. pp. 278--296 (2020)

\bibitem{polymorphic-vampire}
Bhayat, A., Reger, G.: {A Polymorphic \vampire (Short Paper)}. In: IJCAR. pp.
  361--368 (2020)

\bibitem{DBLP:conf/cade/BhayatSR23}
Bhayat, A., Schoisswohl, J., Rawson, M.: {Superposition with Delayed
  Unification}. In: CADE. pp. 23--40 (2023)

\bibitem{DBLP:conf/ijcar/BhayatS24}
Bhayat, A., Suda, M.: {A Higher-Order \vampire (Short Paper)}. In: IJCAR. pp.
  75--85 (2024)

\bibitem{cadical}
Biere, A., Faller, T., Fazekas, K., Fleury, M., Froleyks, N., Pollitt, F.:
  Cadical 2.0. In: CAV. pp. 133--152 (2024).
  \doi{10.1007/978-3-031-65627-9\_7},
  \url{https://doi.org/10.1007/978-3-031-65627-9\_7}

\bibitem{tff1}
Blanchette, J.C., Paskevich, A.: {{TFF1:} The {TPTP} Typed First-Order Form
  with Rank-1 Polymorphism}. In: CADE. pp. 414--420 (2013)

\bibitem{HipSpec}
Claessen, K., Johansson, M., Rosén, D., Smallbone, N.: {Automating Inductive
  Proofs Using Theory Exploration}. In: CADE (2013)

\bibitem{claessen2003new}
Claessen, K., S{\"o}rensson, N.: {New Techniques that Improve {MACE}-style
  Model Finding}. In: WS on Model Computation - Principles, Algorithms and
  Applications (2003)

\bibitem{SATSubsumption24}
Coutelier, R., Rath, J., Rawson, M., Biere, A., Kov\'acs, L.: {SAT Solving for
  Variants of First-Order Subsumption}. Formal Methods in System Design  (2024)

\bibitem{Z3}
De~Moura, L., Bj{\o}rner, N.: {Z3: An Efficient SMT Solver}. In: TACAS. pp.
  337--340 (2008)

\bibitem{sledgehammer}
Desharnais, M., Vukmirovi\'{c}, P., Blanchette, J., Wenzel, M.: {Seventeen
  Provers Under the Hammer}. In: ITP. pp. pp. 8:1--8:18 (2022)

\bibitem{DBLP:conf/sat/EenS03}
E{\'{e}}n, N., S{\"{o}}rensson, N.: {An Extensible {SAT}-solver}. In: SAT. pp.
  502--518 (2003)

\bibitem{substitution-tree}
Graf, P.: {Substitution Tree Indexing}. In: RTA. pp. 117--131 (1995)

\bibitem{gmp}
Granlund, T.: {The GNU Multiple Precision Arithmetic Library} (2023),
  \url{https://gmplib.org/gmp-man-6.3.0.pdf}

\bibitem{tod}
Hajdu, M., Coutelier, R., Kov\'acs, L., Voronkov, A.: {Term Ordering Diagrams}.
  In: CADE (2025), to appear

\bibitem{GettingSaturated22}
Hajdu, M., Hozzov{\'{a}}, P., Kov{\'{a}}cs, L., Reger, G., Voronkov, A.:
  {Getting Saturated with Induction}. In: Principles of Systems Design. pp.
  306--322 (2022)

\bibitem{CICM20}
Hajdu, M., Hozzov{\'a}, P., Kov{\'a}cs, L., Schoisswohl, J., Voronkov, A.:
  {Induction with Generalization in Superposition Reasoning}. In: CICM. pp.
  123--137 (2020)

\bibitem{PAAR22}
Hajdu, M., Kovacs, L., Rawson, M., Voronkov, A.: {The \vampire Approach to
  Induction}. EasyChair Preprint no. 9217 (EasyChair, 2022)

\bibitem{LPAR24Hajdu}
Hajdu, M., Kovács, L., Rawson, M.: {Rewriting and Inductive Reasoning}. In:
  LPAR. pp. 278--294 (2024)

\bibitem{FMCAD21}
Hajdu, M., Hozzová, P., Kovács, L., Voronkov, A.: {Induction with Recursive
  Definitions in Superposition}. In: FMCAD. pp. 1--10 (2021)

\bibitem{IJCAR24}
Hozzov\'{a}, P., Amrollahi, D., Hajdu, M., Kov\'{a}cs, L., Voronkov, A.,
  Wagner, E.M.: {Synthesis of Recursive Programs in Saturation}. In: IJCAR. p.
  154–171 (2024)

\bibitem{CADE23}
Hozzov{\'a}, P., Kov{\'a}cs, L., Norman, C., Voronkov, A.: Program synthesis in
  saturation. In: CADE. pp. 307--324 (2023)

\bibitem{CADE21}
Hozzov{\'a}, P., Kov{\'a}cs, L., Voronkov, A.: {Integer Induction in
  Saturation}. In: CADE. pp. 361--377 (2021)

\bibitem{cpp17}
ISO: ISO/IEC 14882:2017: {Programming} languages --- \CC. International
  Organization for Standardization, Geneva, Switzerland (Dec 2017)

\bibitem{blocked-clause-elimination}
J{\"{a}}rvisalo, M., Biere, A., Heule, M.: {Blocked Clause Elimination}. In:
  TACAS. pp. 129--144 (2010)

\bibitem{CryptoVampire}
Jeanteur, S., Kov{\'{a}}cs, L., Maffei, M., Rawson, M.: {CryptoVampire:
  Automated Reasoning for the Complete Symbolic Attacker Cryptographic Model}.
  In: SP. pp. 3165--3183 (2024)

\bibitem{ACL2book}
Kaufmann, M., Manolios, P., Moore, J.S.: Computer-Aided Reasoning: An Approach,
  vol.~3. Springer (06 2000). \doi{10.1007/978-1-4615-4449-4}

\bibitem{bce}
Kiesl, B., Suda, M., Seidl, M., Tompits, H., Biere, A.: {Blocked Clauses in
  First-Order Logic}. In: LPAR. pp. 31--48 (2017)

\bibitem{cmake}
Kitware, I.: {CMake} (2025), \url{https://cmake.org/}

\bibitem{iProver}
Korovin, K.: {{iProver} --- An Instantiation-Based Theorem Prover for
  First-Order Logic (System Description)}. In: IJCAR. pp. 292--298 (2008)

\bibitem{ALASCA}
Korovin, K., Kov{\'{a}}cs, L., Reger, G., Schoisswohl, J., Voronkov, A.:
  {{ALASCA:} Reasoning in Quantified Linear Arithmetic}. In: TACAS. pp.
  647--665 (2023)

\bibitem{DBLP:conf/cpp/KotelnikovKRV16}
Kotelnikov, E., Kov{\'{a}}cs, L., Reger, G., Voronkov, A.: {The \vampire and
  the {FOOL}}. In: CPP. pp. 37--48 (2016)

\bibitem{DBLP:conf/cade/KotelnikovKV18}
Kotelnikov, E., Kov{\'{a}}cs, L., Voronkov, A.: {A FOOLish Encoding of the Next
  State Relations of Imperative Programs}. In: IJCAR. pp. 405--421 (2018)

\bibitem{IJCAR24InductionInvited}
Kov{\'a}cs, L., Hozzov{\'a}, P., Hajdu, M., Voronkov, A.: {Induction in
  Saturation}. In: IJCAR. pp. 21--29 (2024)

\bibitem{CAV13}
Kovács, L., Voronkov, A.: {First-Order Theorem Proving and \vampire}. In: CAV.
  pp. 1--35 (2013)

\bibitem{CLINGO}
Lifschitz, V., L{\"{u}}hne, P., Schaub, T.: {Towards Verifying Logic Programs
  in the Input Language of clingo}. In: Fields of Logic and Computation {III}.
  pp. 190--209 (2020)

\bibitem{WSL}
Microsoft: {Windows Subsystem for Linux (WSL)},
  \url{https://ubuntu.com/desktop/wsl}

\bibitem{standard-ml}
Milner, R.: {The Definition of Standard ML: Revised}. MIT press (1997)

\bibitem{cygwin}
Racine, J.: {The {Cygwin} Tools: a {GNU} Toolkit for {Windows}} (2000)

\bibitem{term-indexing}
Ramakrishnan, I.V., Sekar, R., Voronkov, A.: {Term Indexing}. In: Handbook of
  Automated Reasoning, pp. 1853--1964. Elsevier and {MIT} Press (2001)

\bibitem{DBLP:conf/gcai/RegerB0V16}
Reger, G., Bj{\o}rner, N.S., Suda, M., Voronkov, A.: {{AVATAR} Modulo
  Theories}. In: {GCAI}. pp. 39--52 (2016)

\bibitem{TheorySimplifications}
Reger, G., Schoisswohl, J., Voronkov, A.: {Making Theory Reasoning Simpler}.
  In: TACAS. pp. 164--180 (2021)

\bibitem{DBLP:conf/sat/Reger0V16}
Reger, G., Suda, M., Voronkov, A.: {Finding Finite Models in Multi-sorted
  First-Order Logic}. In: SAT. pp. 323--341 (2016)

\bibitem{newcnf}
Reger, G., Suda, M., Voronkov, A.: {New Techniques in Clausal Form Generation}.
  In: {GCAI}. pp. 11--23 (2016)

\bibitem{Reger0V18}
Reger, G., Suda, M., Voronkov, A.: {Unification with Abstraction and Theory
  Instantiation in Saturation-Based Reasoning}. In: TACAS. pp. 3--22 (2018)

\bibitem{DBLP:conf/tacas/Reger0V18}
Reger, G., Suda, M., Voronkov, A.: {Unification with Abstraction and Theory
  Instantiation in Saturation-Based Reasoning}. In: TACAS. pp. 3--22 (2018)

\bibitem{CADE19}
Reger, G., Voronkov, A.: {Induction in Saturation-Based Proof Search}. In:
  CADE. pp. 477--494 (2019)

\bibitem{PartiallyAdaptiveCodeTrees}
Riazanov, A., Voronkov, A.: {Partially Adaptive Code Trees}. In: JELIA. pp.
  209--223 (2000)

\bibitem{DBLP:conf/cav/Rungta22}
Rungta, N.: {A Billion {SMT} Queries a Day (Invited Paper)}. In: CAV. pp. 3--18
  (2022)

\bibitem{DBLP:conf/lpar/SchoisswohlKK24}
Schoisswohl, J., Kov{\'{a}}cs, L., Korovin, K.: {{VIRAS:} Conflict-Driven
  Quantifier Elimination for Integer-Real Arithmetic}. In: {LPAR}. pp. 147--164
  (2024)

\bibitem{E19}
Schulz, S., Cruanes, S., Vukmirovi{\'c}, P.: {Faster, Higher, Stronger: {E}
  2.3}. In: CADE. pp. 495--507 (2019)

\bibitem{twee}
Smallbone, N.: {Twee: An Equational Theorem Prover}. In: CADE. pp. 602--613
  (2021)

\bibitem{Zeno}
Sonnex, W., Drossopoulou, S., Eisenbach, S.: {Zeno: An Automated Prover for
  Properties of Recursive Data Structures}. In: TACAS. pp. 407--421 (2012)

\bibitem{leo3}
Steen, A., Benzm{\"{u}}ller, C.: {The Higher-Order Prover {Leo-III}}. In:
  IJCAR. pp. 108--116 (2018)

\bibitem{DBLP:conf/cade/Suda22}
Suda, M.: {\vampire Getting Noisy: Will Random Bits Help Conquer Chaos? (System
  Description)}. In: IJCAR. pp. 659--667 (2022)

\bibitem{Sut16}
Sutcliffe, G.: {The CADE ATP System Competition - CASC}. AI Magazine
  \textbf{37}(2),  99--101 (2016)

\bibitem{TPTP}
Sutcliffe, G.: {The Logic Languages of the TPTP World}. Logic Journal of the
  IGPL  (2022). \doi{10.1093/jigpal/jzac068}

\bibitem{CASC-J12}
Sutcliffe, G.: {The 12th {IJCAR} Automated Theorem Proving System Competition
  --- {CASC-J12}}. The European Journal on Artificial Intelligence
  \textbf{0}(0),  30504554241305110 (0)

\bibitem{DBLP:conf/lpar/Sutcliffe08}
Sutcliffe, G.: {{The {SZS} Ontologies for Automated Reasoning Software}}. In:
  LPAR Workshops (2008)

\bibitem{DBLP:conf/lpar/Sutcliffe10}
Sutcliffe, G.: {The {TPTP} World --- Infrastructure for Automated Reasoning}.
  In: LPAR. pp. 1--12 (2010)

\bibitem{cosmopolitan}
Tunney, J.: Cosmopolitan {Libc} (2025), \url{https://justine.lol/cosmopolitan/}

\bibitem{DBLP:conf/cav/Voronkov14}
Voronkov, A.: {{AVATAR:} The Architecture for First-Order Theorem Provers}. In:
  CAV. pp. 696--710 (2014)

\bibitem{Spider}
Voronkov, A.: {{Spider}: Learning in the Sea of Options} (2023),
  \url{https://easychair.org/smart-program/Vampire23/2023-07-05.html#talk:223833}

\bibitem{zipperposition}
Vukmirovic, P., Bentkamp, A., Blanchette, J., Cruanes, S., Nummelin, V.,
  Tourret, S.: {Making Higher-Order Superposition Work}. J. Autom. Reason.
  \textbf{66}(4),  541--564 (2022)

\bibitem{WCDHNR19}
Weber, T., Conchon, S., D{\'{e}}harbe, D., Heizmann, M., Niemetz, A., Reger,
  G.: The {SMT} competition 2015-2018. J. Satisf. Boolean Model. Comput.
  \textbf{11}(1),  221--259 (2019), \url{https://doi.org/10.3233/SAT190123}

\bibitem{Spass09}
Weidenbach, C., Dimova, D., Fietzke, A., Kumar, R., Suda, M., Wischnewski, P.:
  {SPASS Version 3.5}. In: {CADE}. pp. 140--145 (2009)

\end{thebibliography}
